\documentclass[12pt,a4paper]{article}
\usepackage[utf8]{inputenc}
\usepackage{epsfig}
\usepackage{mathtools}
\usepackage{amsfonts}
\usepackage{amssymb}
\usepackage{amsmath}
\usepackage{cancel}
\usepackage{color}
\usepackage{slashed}
\usepackage{cite}
\usepackage{caption}
\usepackage{subcaption}
\usepackage{comment}
\usepackage{marginnote}
\usepackage{accents}

\usepackage{geometry} 
\usepackage{graphicx}

\usepackage{multirow}           
\usepackage{array}

\allowdisplaybreaks

\newcommand{\ep}{\varepsilon}
\newcommand{\OO}{\mathcal{O}}

\def\ep  {\varepsilon}

\begin{document}

\begin{center}
	
\vspace{2cm}		
		
{\Large\bf On series and integral representations \\ of some NRQCD master integrals} \vspace{1cm}

{\large M.A. Bezuglov$^{1,2,3}$, A.V. Kotikov$^{1}$, A.I. Onishchenko$^{1,3,4}$}\vspace{0.5cm}
		
{\it
$^1$Bogoliubov Laboratory of Theoretical Physics,  Joint
Institute for Nuclear Research,\\ Dubna, Russia, \\
$^2$Moscow Institute of Physics and Technology (State University), Dolgoprudny, Russia, \\
$^3$Budker Institute of Nuclear Physics, Novosibirsk, Russia, \\
$^4$Skobeltsyn Institute of Nuclear Physics,  Moscow State University, Moscow, Russia}
\vspace{1cm}

\abstract{We consider new ways of obtaining series and integral representations for master integrals arising in the process of matching of QCD to NRQCD. The latter results are exact in space-time dimension $d$. In addition, we discuss series expansion of the obtained results at fixed values of $d$.
}

\end{center}

\newpage

\section{Introduction}

At present we have a lot of techniques for calculating multiloop Feynman diagrams, see \cite{Weinzierl-review,Duhr-review,Kotikov-review} for recent reviews. The available techniques could be separated into two wide classes, such as the solution of some system of equations, in particular the system of differential equations\cite{diffeqn1,diffeqn2,diffeqn3,diffeqn4,diffeqn5,epform1,epform2}
or direct integration of parametric representations \cite{BrownModuli,linear-reducibility-1,linear-reducibility-2,PanzerAlgorithms,directint1,directint2,directint3}. In many cases the results for Feynman diagrams can be written in terms of multiple polylogarithms (MPLs) \cite{polylog1,polylog2,polylog3}, which is well studied class of functions at the moment. When the problem solves in terms of MPLs it is either due to the fact that the corresponding differential system reduces to the so called $\ep$-form \cite{epform1,epform2,epform-criterium} or corresponding parametric representation has the property of linear reducibility \cite{linear-reducibility-1,linear-reducibility-2}. In other cases we are required to introduce both new functions and techniques. For example, in the context of differential equation method besides a new class of functions one may either allow for non-algebraic transformation to $\ep$-form \cite{epform-elliptics} or use a notion of regular basis for non-polylogarithmic integrals \cite{ep-regular-basis}. The direct integration algorithms also require extension when going beyond MPLs \cite{Broedel:2017kkb,Broedel:2017siw,Broedel:2019hyg,LinearReducibledEllipticFeynmanIntegrals}. The first simplest functions which one encounters beyond MPLs are the so-called elliptic polylogarithms (EPLs) \cite{Beilinson:1994,Wildeshaus,Levin:1997,Levin:2007,Enriquez:2010,Brown:2011,Bloch:2013tra,Adams:2014vja,Bloch:2014qca,Adams:2015gva,Adams:2015ydq,Adams:2016xah,Remiddi:2017har,Broedel:2017kkb,Broedel:2017siw,Broedel:2018iwv,Broedel:2018qkq,Broedel:2019hyg,Broedel:2019tlz,Bogner:2019lfa,Broedel:2019kmn,Walden:2020odh,Weinzierl:2020fyx,kites-elliptic}.
Next, come problems with several elliptic curves \cite{Adams:2018bsn,Adams:2018kez}
and completely new functions, such as in \cite{Bloch:2014qca,Primo:2017ipr,Bourjaily:2017bsb,Bourjaily:2018ycu,Bourjaily:2018yfy,kites-elliptic,Bonisch:2021yfw}.

In the present short note, we use an example a set of two-loop master integrals arising in the process of matching of QCD to NRQCD to introduce several new methods for obtaining their series and integral representations, which are either exact in the value of space-time dimension or expanded to any prescribed order for its fixed values. 
The mentioned techniques are analytical Frobenius method\footnote{For previous applications of Frobenius method in the context of Feynman diagrams see for example \cite{Frobenius1,Frobenius2,Frobenius3,Frobenius4,Frobenius5,KKOVelliptic2,Bonisch:2021yfw}} as introduced in \cite{nonplanarVertex}, the use of Feynman parameter trick\footnote{Similar trick was used before in \cite{effective-mass1,effective-mass2,effective-mass3,effective-mass4} under the name of effective mass approach \cite{diffeqn1,diffeqn3,effective-mass5}.} and differential equations with respect to the latter \cite{KKOVelliptic1,KKOVelliptic2,LinearReducibledEllipticFeynmanIntegrals,kites-elliptic} and integral representations for hypergeometric functions \cite{KKOVelliptic1,KKOVelliptic2,kotikov-EMPL}.

This note has the following structure. In the next section we consider
analytical Frobenius solutions for general values of space-time dimension. Section \ref{expandedIntegralReps} describes the use of Feynman parameter trick to reduce the problem of evaluation of two-loop diagrams to effective one-loop diagrams and the use of differential equations with respect to mentioned Feynman parameter for the calculation of the latter. Next, in section \ref{exactIntegralReps} we show how the exact Frobenius results in terms of hypergeometric $~_p F_q$-functions can be transformed into corresponding integral representations exact in space-time dimension. Finally, section \ref{conclusion} contains our conclusion. Appendix \ref{II-notation} contains notation for iterated integrals with algebraic kernels used to present some of our results.

\section{Analytical Frobenius solution}
\label{analyticalFrobenius}

\begin{figure}[h]
	\center{\includegraphics[width=0.6\textwidth]{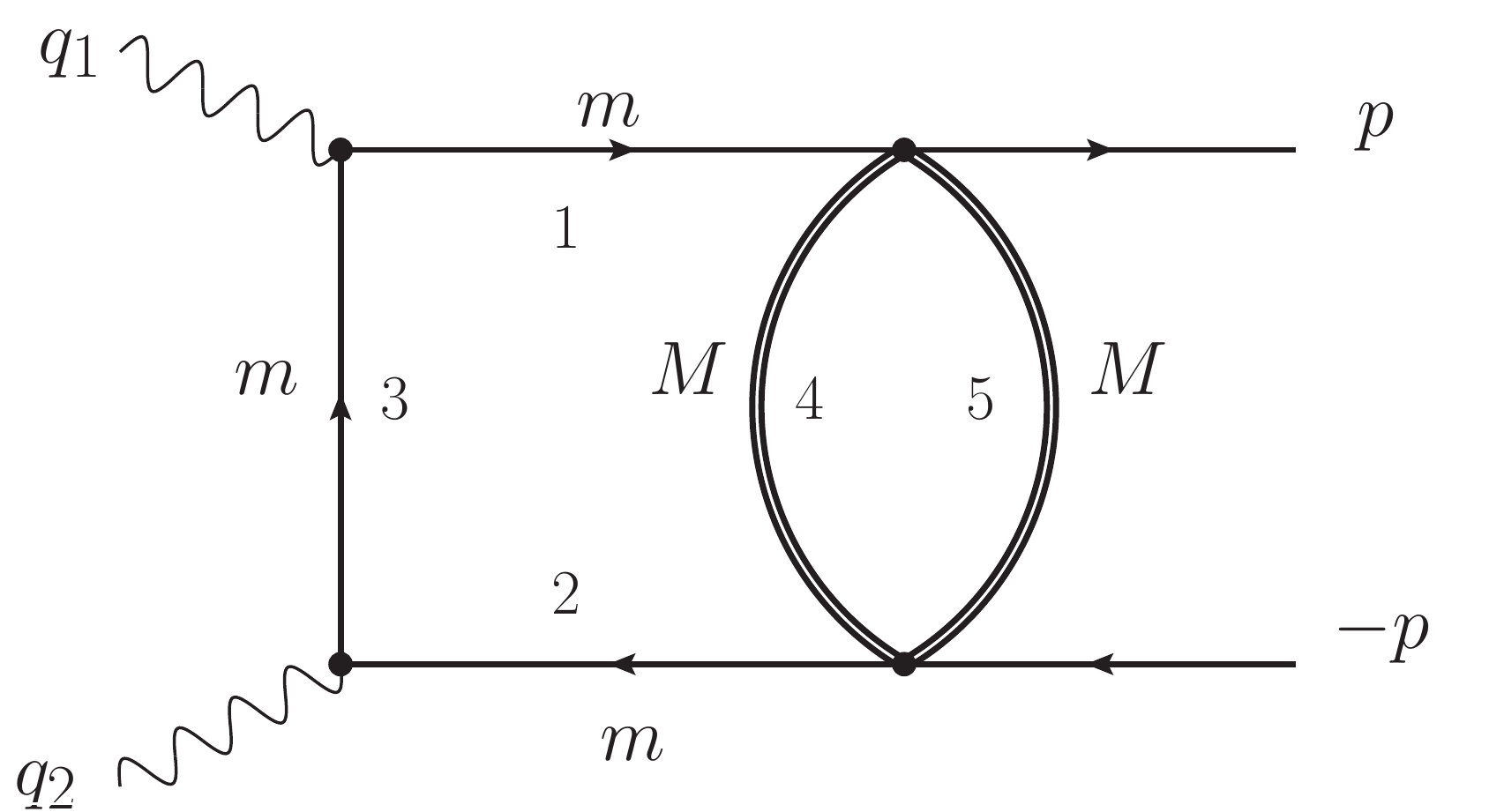}}
	\caption{Graphical representation for the family of integrals in Eq. \eqref{eq:2loopFamily}.}
	\label{fig:familygraph}
\end{figure}

We will be interested in master integrals for a family of Feynman integrals considered previously in \cite{KKOVelliptic1,KKOVelliptic2}. The latter is defined as 
\begin{multline}
J_{b_1 b_2 b_3 a_1 a_2}^{mM} = \int\frac{d^d k d^d l }{\pi^2}
\frac{1}{((k+q_1)^2-m^2)^{b_1}((k-q_2)^2-m^2)^{b_2}(k^2-m^2)^{b_3}} \\
\times \frac{1}{((l+\frac{q_1-q_2}{2})^2-M^2)^{a_1}((l-k)^2-M^2)^{a_2}}\, , \label{eq:2loopFamily}
\end{multline}
where $q_1^2=q_2^2=0$ and $q_1\cdot q_2 = 2m^2$.
A graphical representation of these family of integrals can be found in Fig. \ref{fig:familygraph} where we define $p = \frac{1}{2}\left(q_1+q_1\right)$.
It is convenient to set $M=1$ and $m^2=x$, while the full dependence on $m$ and $M$ can be easily restored if required.  Using integration by parts (IBP)  relations \cite{IBP1,IBP2} all integrals in this family can be reduced to the set of 9 master integrals. The latter can be chosen as 
\begin{align}
(J_1,\ldots , J_9)^\top =& \Big\{
J^{mM}_{00011}, J^{mM}_{00101}, J^{mM}_{00111}, J^{mM}_{00211}, J^{mM}_{00121}, J^{mM}_{01011}, J^{mM}_{02011}, J^{mM}_{01111}, J^{mM}_{11101}
 \Big\}^\top
\end{align} 
Three of these master integrals can be easily computed using direct integration in Feynman parameters and we have
\begin{equation}
J_1 = -\Gamma (1-\frac{d}{2})^2 , J_2 = -x^{\frac{d}{2}-1}\Gamma (1-\frac{d}{2})^2 , J_9 =\frac{(d-2)\Gamma (1-\frac{d}{2})^2 x^{\frac{d}{2}-3}}{8}\left[
\Psi (\frac{1}{2}) - \Psi (\frac{d-3}{2})
\right],
\end{equation}
To calculate the rest we will employ analytical Frobenius solution\footnote{For previous applications of Frobenius method in the context of Feynman diagrams see for example \cite{Frobenius1,Frobenius2,Frobenius3,Frobenius4,Frobenius5,KKOVelliptic2,Bonisch:2021yfw}} of their differential equations as proposed in \cite{nonplanarVertex}. For example, using IBP relations it is not hard to find that $J_3$ master integral satisfies the following differential equation
\begin{multline}
(4+x^2)x^3\frac{d^3 J_3}{dx^3}+(16-2d+12x^2-\frac{5d}{2}x^2)x^2\frac{d^2 J_3}{dx^2}+(3+d(2-d)+(d-4)(2d-9)x^2)x\frac{d J_3}{dx} \\
+ \frac{1}{2}(11d-6-d^2(6-d)-(d-4)^2(d-3x^2))J_3
+\frac{1}{2}(d-2)^2J_2 -\frac{1}{4}(d-2)^2(d-1+(d-3)x)J_1 = 0
\label{eq:J3eqn}
\end{multline}
Noting that the coefficients of this equation in front of $J_3$, $x^3\frac{d^3 J_3}{dx^3}$, $x^2\frac{d^2 J_3}{dx^2}$ and $x\frac{d J_3}{dx}$ are of the form $a+bx^2$ it is natural to look for the power series solution in the form of the ansatz
\begin{equation}
\label{eq:J3GeneralSer}
J_3 = \sum_{n=0}^{\infty} c_n^{\lambda} x^{2n+\lambda}
\end{equation} 
Substituting the latter into Eq.\eqref{eq:J3eqn} and equating coefficients in front of the same powers of $x$ we get the recurrence relation for the coefficients $c_n^{\lambda}$:
\begin{multline}
(d-3-4n-2\lambda)(d-1-4n-2\lambda)(d-2+4n+2\lambda)c_n^{\lambda} \\
- (d-4n-2\lambda)(d-2-2n-\lambda)(d-1-2n-\lambda)c_{n-1}^{\lambda} = 0 \label{eq:J3cn-recurrence}
\end{multline}
The solution of this first order difference equation is easy and we obtain
\begin{equation}
c_n^{\lambda} = \frac{(-1)^n}{4^n}\frac{\Gamma (1-\frac{d}{4}+n+\frac{\lambda}{2})\Gamma (3-d+2n+\lambda)}{\Gamma (\frac{5}{2}-\frac{d}{2}+2n+\lambda)\Gamma (\frac{1}{2}+\frac{d}{4}+n + \frac{\lambda}{2})} C_{\lambda}\, ,
\end{equation}
where $C_{\lambda}$ is some constant. The $\lambda$'s corresponding to homogeneous solutions are determined from Eq.\eqref{eq:J3cn-recurrence} at $n=0$ with the requirement that $c_{-1}^{\lambda} = 0$. This way we get
\begin{equation}
\lambda_{hom}\in \left\{
\frac{2-d}{2}, \frac{d-3}{2}, \frac{d-1}{2}
\right\}\, .
\end{equation}
Then the general solution with account of particular solutions corresponding to inhomogeneities proportional to $J_1$ and $J_2$ is given by
\begin{equation}
J_3 = \sum_{\lambda\in \{0,1,\frac{d-2}{2}, \frac{2-d}{2}, \frac{d-3}{2}, \frac{d-1}{2} \}}\sum_{n=0}^{\infty} c_n^{\lambda} x^{2n+\lambda}
\end{equation}  
The constants $C_{\bullet}$ for particular solutions of nonhomogeneous equation are found by substituting this solution into Eq.\eqref{eq:J3eqn} and requiring that coefficients in front of $x^0$, $x^1$ and $x^{d/2-1}$ are zero. Next, constants $C_{\bullet}$ for homogeneous solutions are determined from boundary conditions at $x=0$. It turns out that they all zero and we finally get 
\begin{multline}
J_3 = \frac{\pi \csc (\frac{\pi d}{2})\Gamma (2-\frac{d}{2})}{\Gamma (\frac{d}{2})} \Big\{
x^{-1+\frac{d}{2}} ~_4 F_3 \left(
\begin{array}{c}
1, \frac{1}{2}, \frac{4-d}{4}, \frac{6-d}{4} \\
\frac{3}{4}, \frac{5}{4}, \frac{d}{2}
\end{array} \Big| -\frac{x^2}{4}
\right) \\
+\frac{1}{d-3} ~_4 F_3 \left(
\begin{array}{c}
1, \frac{3-d}{2}, \frac{4-d}{2}, \frac{4-d}{4} \\
\frac{5-d}{4}, \frac{7-d}{4}, \frac{2+d}{4}
\end{array} \Big| -\frac{x^2}{4}
\right)
+ \frac{(d-2)}{d (d-5)} ~_4 F_3 \left(
\begin{array}{c}
1, \frac{4-d}{2}, \frac{5-d}{2}, \frac{6-d}{2} \\
\frac{7-d}{4}, \frac{9-d}{4}, \frac{4+d}{4}
\end{array} \Big| -\frac{x^2}{4}
\right)
\Big\}\, ,
\end{multline}
The expressions for $J_4$ and $J_5$ are expressed thorough $J_3$ and it first and second derivatives. For example, from IBP relations we have
\begin{equation}
J_5 = -\frac{x}{2}\frac{d J_3}{dx} + \frac{(d-3)}{2}J_3\, .
\end{equation}
This way we get
\begin{multline}
J_4 = \frac{\pi^2\csc (\frac{\pi d}{2})^2}{(d-5)\Gamma (\frac{d-4}{2})\Gamma (\frac{d}{2})}\Big\{ \\
-\frac{(d-5)(d-2)x^{\frac{d}{2}-2}}{(d-4)} ~_4 F_3 \left(
\begin{array}{c}
1, \frac{1}{2}, \frac{4-d}{4}, \frac{6-d}{4} \\
\frac{3}{4}, \frac{5}{4}, \frac{d-2}{2}
\end{array} \Big| -\frac{x^2}{4}
\right)
+\frac{(d-5)x^{\frac{d}{2}-1}}{6} ~_4 F_3 \left(
\begin{array}{c}
1,\frac{3}{2}, \frac{6-d}{4}, \frac{8-d}{4} \\
\frac{5}{4}, \frac{7}{4}, \frac{d}{2}
\end{array} \Big| -\frac{x^2}{4}
\right) \\
-\frac{(d-2)}{2(d-4)} ~_4 F_3 \left(
\begin{array}{c}
1, \frac{4-d}{2}, \frac{5-d}{2}, \frac{6-d}{4} \\
\frac{7-d}{4}, \frac{9-d}{4}, \frac{d}{4}
\end{array} \Big| -\frac{x^2}{4}
\right) +\frac{1}{2} ~_4 F_3 \left(
\begin{array}{c}
1, \frac{4-d}{2}, \frac{5-d}{2}, \frac{8-d}{4} \\
\frac{7-d}{4}, \frac{9-d}{4}, \frac{2+d}{4} 
\end{array} \Big| -\frac{x^2}{4}
\right) \\
-\frac{(d-4) x}{2 (d-7)} ~_4 F_3 \left(
\begin{array}{c}
1, \frac{5-d}{2}, \frac{6-d}{2}, \frac{8-d}{4} \\
\frac{9-d}{4}, \frac{11-d}{4}, \frac{2+d}{4}
\end{array} \Big| -\frac{x^2}{4}
\right) + \frac{(d-2)(d-6)x}{2 d (d-7)} ~_4 F_3 \left(
\begin{array}{c}
1, \frac{5-d}{2}, \frac{6-d}{2}, \frac{10-d}{4} \\
\frac{9-d}{4}, \frac{11-d}{4}, \frac{4+d}{4}
\end{array} \Big| -\frac{x^2}{4}
\right)
\Big\}
\end{multline}
and
\begin{multline}
J_5 = -\frac{\pi \csc (\frac{\pi d}{2})\Gamma (3-\frac{d}{2})}{2\Gamma (\frac{d}{2})}\Big\{
x^{-1+\frac{d}{2}} ~_4 F_3 \left(
\begin{array}{c}
1, \frac{1}{2}, \frac{6-d}{4}, \frac{8-d}{4} \\
\frac{3}{4}, \frac{5}{4}, \frac{d}{2}
\end{array} \Big| -\frac{x^2}{4}
\right) \\
+ \frac{2}{d-4} ~_4 F_3 \left(
\begin{array}{c}
1, \frac{4-d}{2}, \frac{5-d}{2}, \frac{4-d}{4} \\
\frac{5-d}{4}, \frac{7-d}{4}, \frac{2+d}{4} 
\end{array} \Big| -\frac{x^2}{4}
\right) + \frac{2 (d-2) t}{d (d-5)} ~_4 F_3 \left(
\begin{array}{c}
1, \frac{5-d}{2}, \frac{6-d}{2}, \frac{6-d}{4} \\
\frac{7-d}{4}, \frac{9-d}{4}, \frac{4+d}{4}
\end{array} \Big| -\frac{x^2}{4}
\right) 
\Big\}\, .
\end{multline}
The expressions for $J_6$ and $J_7$ master integrals are obtained along the same lines and we have
\begin{multline}
J_6 = -\frac{\pi \csc (\frac{\pi d}{2})\Gamma (2-\frac{d}{2})}{\Gamma (4-d)\Gamma (\frac{d}{2})} 
 \Big\{
\Gamma (3-d) ~_3 F_2 \left(
\begin{array}{c}
1, \frac{1}{2}, 3-d \\
\frac{5-d}{2}, \frac{d}{2}
\end{array} \Big| x
\right) - x^{-1+\frac{d}{2}} \Gamma (4-d)
~_3 F_2 \left(
\begin{array}{c}
\frac{4-d}{2}, 1, \frac{d-1}{2} \\
\frac{3}{2}, d-1
\end{array} \Big| x
\right)
\Big\}\, ,
\end{multline}
and
\begin{multline}
J_7 = \frac{\pi^2 \csc (\frac{\pi d}{2})^2}{2\Gamma (\frac{d-2}{2})\Gamma (\frac{d}{2})}   \Big\{
\frac{1}{5-d} ~_3 F_2 \left(
\begin{array}{c}
\frac{3}{2}, 1, 4-d \\
\frac{7-d}{2}, \frac{d}{2}
\end{array} \Big| x
\right) - (d-2) x^{-2+\frac{d}{2}}
~_3 F_2 \left(
\begin{array}{c}
\frac{4-d}{2}, 1, \frac{d-1}{2} \\
\frac{3}{2}, d-2
\end{array} \Big| x
\right)
\Big\}\, .
\end{multline}
Finally, the expression for $J_8$ master integral is found with the use of its differential equation 
\begin{equation}
x^{\frac{d}{2}-1}\frac{d}{dx}\left(
x^{\frac{6-d}{2}}J_8 
\right) - x J_7 - J_5 - x J_4  + \frac{3d-8}{4} J_3 - \frac{(d-2)^2}{8(d-3)}J_1 = 0
\end{equation}
The latter is easily integrated and we get
\begin{multline}
J_8 = \frac{\pi^2 \csc (\frac{\pi d}{2})^2}{\Gamma (\frac{d-2}{2})^2} \Bigg\{
\frac{1}{(d-3)(d-4)x} ~_4 F_3 \left(
\begin{array}{c}
\frac{1}{2}, 1, 3-d, \frac{4-d}{2} \\
\frac{6-d}{2}, \frac{5-d}{2}, \frac{d-2}{2}
\end{array} \Big| x
\right) \\
-x^{\frac{d}{2}-2} ~_4 F_3 \left(
\begin{array}{c}
1, 1, \frac{4-d}{2}, \frac{d-1}{2} \\
2, \frac{3}{2}, d-2
\end{array} \Big| x
\right)
+ \frac{(d-4)x^{-1+\frac{d}{2}}}{12 (d-2)} ~_5 F_4 \left(
\begin{array}{c}
1, 1, \frac{3}{2}, \frac{6-d}{4}, \frac{8-d}{4} \\
2, \frac{5}{4}, \frac{7}{4}, \frac{d}{2}
\end{array} \Big| -\frac{x^2}{4}
\right) \\
-\frac{1}{(d-3)(d-4)x} ~_5 F_4 \left(
\begin{array}{c}
1, \frac{3-d}{2}, \frac{4-d}{2}, \frac{4-d}{4}, \frac{6-d}{4} \\
\frac{8-d}{4}, \frac{5-d}{4}, \frac{7-d}{4}, \frac{d}{4}
\end{array} \Big| -\frac{x^2}{4}
\right) \\
-\frac{(d-4)}{(d-2)(d-5)(d-6)} ~_5 F_4\left(
\begin{array}{c}
1, \frac{4-d}{2}, \frac{5-d}{2}, \frac{6-d}{4}, \frac{8-d}{4} \\
\frac{10-d}{4}, \frac{7-d}{4}, \frac{9-d}{4}, \frac{2+d}{4}
\end{array} \Big| -\frac{x^2}{4}
\right)
\Bigg\}\, . \label{eq:J8-hypergeometry}
\end{multline}

\section{Integral representations expanded at fixed values  of $d$}
\label{expandedIntegralReps}

Before discussing integral representation of our master integrals for general space-time dimension in next section lets see how we can get integral representations expanded at fixed values of space-time dimension. To be specific let us consider $\ep$ expansion at $d=2-2\ep$. The starting point in getting such type of integral representation is the relation between two and one loop master integrals considered in \cite{KKOVelliptic1,KKOVelliptic2,kites-elliptic}. For $J_{0b_2b_311}^{mM}$ master integrals\footnote{$J^{mM}_{00121}$ master integral can be expressed via new $J^{mM}_{00311}$ master integral using IBP relations} using Feynman parameters for two last propagators and performing integration over $l$ momentum we get ($x=y^2z, y^2=\frac{1}{s\bar{s}}=\frac{1}{s(1-s)}$):
\begin{multline}
J_{0b_2b_311}^{mM} = (-1)^{b_2+b_3+1}\Gamma (2\ep) \int_0^1 ds (y^2)^{1-\ep-b_2-b_3} \\\times \int\frac{d^{2-2\ep} k}{i\pi^{1-\ep}}\frac{1}{(z-(k-\tilde{q}_2)^2)^{b_2}(z-k^2)^{b_3}(1-(k+(\tilde{q}_1-\tilde{q}_2)/2)^2)^{1+\ep}}\, ,
\end{multline}
where $\tilde{q}_1^2=\tilde{q}_2^2=0$, $\tilde{q}_1\cdot \tilde{q}_2 = 2z$. Or using variable change ($t=\frac{4}{y^2}$)
\begin{equation}
J_{0b_2b_311}^{mM} = \frac{(-1)^{b_2+b_3+1}}{2}\Gamma (2\ep) \int_0^1 \frac{dt}{\sqrt{1-t}} \left(
\frac{4}{t}
\right)^{1-\ep-b_2-b_3} J_{b_2b_31}^{(1)}\, , \label{eq:2loop-1loop-relation}
\end{equation} 
where 
\begin{equation}
J_{a_1 a_2 a_3}^{(1)} = \int\frac{d^{2-2\ep} k}{i\pi^{1-\ep}}
\frac{1}{(z-(k-\tilde{q}_2)^2)^{a_1}(z-k^2)^{a_2}(1-(k+(\tilde{q}_1-\tilde{q}_2)/2)^2)^{a_3+\ep}}
\end{equation}
and $z=\frac{xt}{4}$. Next, using IBP relations all one-loop integrals $J_{a_1 a_2 a_3}^{(1)}$ can be reduced to the following 5 master integrals written as a vector
\begin{equation}
{\bf J} = (J_{000}^{(1)}, J_{010}^{(1)}, J_{020}^{(1)}, J_{100}^{(1)}, J_{200}^{(1)})^\top
\end{equation} 
It is convenient to consider two subsets of these master integrals ${\bf J}_1 = (J_{000}^{(1)}, J_{010}^{(1)}, J_{020}^{(1)})^\top$ and ${\bf J}_2 =(J_{000}^{(1)},  J_{100}^{(1)}, J_{200}^{(1)})^\top$ separately as their differential systems decouple. Consider as an example the first set. The original differential equations system for masters $J_{000}^{(1)}$, $J_{010}^{(1)}$ and $J_{020}^{(0)}$ using balance transformations \cite{epform2} can be further reduced to $\ep$-form \cite{epform1,epform2,epform-criterium}:
\begin{equation}
\frac{d\widetilde{\bf J}_1}{dz} = \ep {\bf M}\widetilde{\bf J}_1
\end{equation}
where ($z_1=\sqrt{1+4z^2}$) 
\begin{equation}
{\bf M} = \left(
\begin{array}{ccc}
0 & 0 & 0 \\
0 & -\frac{2}{z} & \frac{6 z}{z_1}-\frac{3 z_1}{2 z} \\
\frac{2 z}{z_1}-\frac{2}{z_1}-\frac{z_1}{2 z} & \frac{2 z_1}{z}-\frac{8 z}{z_1} & \frac{2}{z}-\frac{16
	z}{z_1^2} \\
\end{array}
\right)
\end{equation}
and $\widetilde{\bf J}_1 = (\widetilde{J}_{1,1}, \widetilde{J}_{1,2}, \widetilde{J}_{1,3})^\top$ is the so called canonical basis. The transformation matrix to the latter (${\bf J}_1 = {\bf T}\widetilde{\bf J}_1$) is given by
\begin{equation}
{\bf T} = \left(
\begin{array}{ccc}
1 & 0 & 0 \\
0 & \frac{4 (2 \epsilon -1)}{3 \epsilon } & 0 \\
\frac{2 \epsilon -1}{2 z} & \frac{2 (2 \epsilon -1)}{z} & \frac{(2 z+1) (2 \epsilon -1)}{z z_1} \\
\end{array}
\right)\, .
\end{equation}
The boundary conditions for canonical master integrals are given by
\begin{equation}
\widetilde{\bf J}_1\Big|_{z\to 0}\sim {\bf L}.{\bf C}\, ,
\end{equation}
where 
\begin{equation}
{\bf L} = \left(
\begin{array}{ccc}
1 & 0 & 0 \\
-\frac{3}{4} & \frac{3 \epsilon }{4 (2 \epsilon -1)} & \frac{3 \epsilon }{4 (2 \epsilon -1)} \\
1 & -\frac{\epsilon }{2 (2 \epsilon -1)} & -\frac{3 \epsilon }{2 (2 \epsilon -1)} \\
\end{array}
\right)
\end{equation}
and 
\begin{equation}
{\bf C} = (c_1^0, c_2^{-\ep}, c_2^{\ep})^\top
\end{equation}
Here $c_i^j$ is the coefficient in front of $z^j$ power in the samll $z$ expansion of $i$-th integral in vector of master integrals ${\bf J}_1$. For these constants we have
\begin{equation}
c_1^0 = \frac{\Gamma (2\ep - 1)}{\Gamma (\ep)}, \quad c_2^{-\ep} = \Gamma (\ep) ,\quad c_2^\ep = 0.
\end{equation}
Now, the solution of differential equations system is easy and we have\footnote{See Appendix \ref{II-notation} for our notation for iterated integrals $\text{II}$'s.}:
\begin{multline}
J_{010}^{(1)} = 
\frac{1}{2 \epsilon }
-\text{II}_8
+
\epsilon  \Big[-16 \text{II}_{\text{2,2}}+4 \text{II}_{\text{2,3}}-8
\text{II}_{\text{2,4}}+4 \text{II}_{\text{3,2}}-\text{II}_{\text{3,3}}+2 \text{II}_{\text{3,4}}+2
\text{II}_{\text{8,8}}-\frac{\pi ^2}{24}\Big] \\
+ 
\epsilon ^2 \Big[\frac{2 \pi ^2 \text{II}_2}{3}+256 \text{II}_{\text{2,1,2}}-64
\text{II}_{\text{2,1,3}}+128 \text{II}_{\text{2,1,4}}+48 \text{II}_{\text{2,2,8}}-12
\text{II}_{\text{2,3,8}}-32 \text{II}_{\text{2,8,2}}+8 \text{II}_{\text{2,8,3}}-16
\text{II}_{\text{2,8,4}} \\ -\frac{\pi ^2 \text{II}_3}{6}-64 \text{II}_{\text{3,1,2}}+16
\text{II}_{\text{3,1,3}}-32 \text{II}_{\text{3,1,4}}-12 \text{II}_{\text{3,2,8}}+3
\text{II}_{\text{3,3,8}}+8 \text{II}_{\text{3,8,2}}-2 \text{II}_{\text{3,8,3}}+4
\text{II}_{\text{3,8,4}}+\frac{\pi ^2 \text{II}_8}{12} \\ +32 \text{II}_{\text{8,2,2}}-8
\text{II}_{\text{8,2,3}}+16 \text{II}_{\text{8,2,4}}-8 \text{II}_{\text{8,3,2}}+2
\text{II}_{\text{8,3,3}}-4 \text{II}_{\text{8,3,4}}-4 \text{II}_{\text{8,8,8}}+\frac{5 \zeta
	(3)}{6}\Big] + \OO (\ep^3) \\
\end{multline} 
and 
\begin{multline}
J_{020}^{(1)} = 
\frac{1}{z}
+\frac{\epsilon}{z}  \Big[-\frac{2 (2 z+1) \text{II}_2}{ z_1}+\frac{(2 z+1) \text{II}_3}{2 
	z_1}-\frac{(2 z+1) \text{II}_4}{ z_1}-\frac{3 \text{II}_8}{2 }\Big]
+
\frac{\epsilon ^2}{z} \Big[
\frac{64 z \text{II}_{\text{1,2}}}{z_1}+\frac{32 \text{II}_{\text{1,2}}}{z_1}-\frac{16 z
	\text{II}_{\text{1,3}}}{z_1}\\ -\frac{8 \text{II}_{\text{1,3}}}{z_1}+\frac{32 z
	\text{II}_{\text{1,4}}}{z_1}+\frac{16 \text{II}_{\text{1,4}}}{z_1}-24 \text{II}_{\text{2,2}}+6
\text{II}_{\text{2,3}}-12 \text{II}_{\text{2,4}}+\frac{12 z \text{II}_{\text{2,8}}}{z_1}+\frac{6
	\text{II}_{\text{2,8}}}{z_1}+6 \text{II}_{\text{3,2}}-\frac{3 \text{II}_{\text{3,3}}}{2}+3
\text{II}_{\text{3,4}}-\frac{3 z \text{II}_{\text{3,8}}}{z_1}\\-\frac{3 \text{II}_{\text{3,8}}}{2
	z_1}-\frac{8 z \text{II}_{\text{8,2}}}{z_1}-\frac{4 \text{II}_{\text{8,2}}}{z_1}+\frac{2 z
	\text{II}_{\text{8,3}}}{z_1}+\frac{\text{II}_{\text{8,3}}}{z_1}-\frac{4 z
	\text{II}_{\text{8,4}}}{z_1}-\frac{2 \text{II}_{\text{8,4}}}{z_1}+3 \text{II}_{\text{8,8}}+\frac{\pi
	^2 z}{6 z_1}+\frac{\pi ^2}{12 z_1}
\Big] + \OO (\ep^3)
\end{multline}
Note, that such solutions can be easily written for any prescribed order in $\ep$. The consideration of the second subset of master integrals ${\bf J}_2$ goes along the same line and we have ($z_2=\sqrt{1-4z}$): 
\begin{multline}
J_{100}^{(1)} = 
\frac{1}{2
	\epsilon }
-\text{II}_8
+\epsilon 
\Big[-\text{II}_{\text{6,6}}-4 \text{II}_{\text{6,7}}-4 \text{II}_{\text{7,6}}-16
\text{II}_{\text{7,7}}+2 \text{II}_{\text{8,8}}-\frac{\pi ^2}{24}\Big]
+
\epsilon ^2 \Big[-\frac{1}{6} \pi ^2 \text{II}_6-8 \text{II}_{\text{6,5,6}}-32
\text{II}_{\text{6,5,7}} \\ +3 \text{II}_{\text{6,6,8}}+12 \text{II}_{\text{6,7,8}}-2
\text{II}_{\text{6,8,6}}-8 \text{II}_{\text{6,8,7}}-\frac{2 \pi ^2 \text{II}_7}{3}-32
\text{II}_{\text{7,5,6}}-128 \text{II}_{\text{7,5,7}}+12 \text{II}_{\text{7,6,8}}+48
\text{II}_{\text{7,7,8}}-8 \text{II}_{\text{7,8,6}}\\ -32 \text{II}_{\text{7,8,7}}+\frac{\pi ^2
	\text{II}_8}{12}+2 \text{II}_{\text{8,6,6}}+8 \text{II}_{\text{8,6,7}}+8 \text{II}_{\text{8,7,6}}+32
\text{II}_{\text{8,7,7}}-4 \text{II}_{\text{8,8,8}}+\frac{5 \zeta (3)}{6}\Big] + \OO (\ep^3)
\end{multline}
and
\begin{multline}
J_{200}^{(1)} = \frac{1}{z}
+\frac{\epsilon}{z} 
\Big[\frac{\text{II}_6}{2 z_2}+\frac{2 \text{II}_7}{z_2}-\frac{3 \text{II}_8}{2}\Big]
+\frac{\epsilon ^2}{z} \Big[\frac{4 \text{II}_{\text{5,6}}}{z_2}+\frac{16 \text{II}_{\text{5,7}}}{z_2}-\frac{3
	\text{II}_{\text{6,6}}}{2}-6 \text{II}_{\text{6,7}}-\frac{3 \text{II}_{\text{6,8}}}{2 z_2}\\ -6
\text{II}_{\text{7,6}}-24 \text{II}_{\text{7,7}} -\frac{6
	\text{II}_{\text{7,8}}}{z_2}+\frac{\text{II}_{\text{8,6}}}{z_2}+\frac{4
	\text{II}_{\text{8,7}}}{z_2}+3 \text{II}_{\text{8,8}}+\frac{\pi ^2}{12 z_2}\Big] + \OO (\ep^3)
\end{multline}
The integrals $J^{(1)}_{a_1 a_2 1}$ entering the expression for two-loop masters \eqref{eq:2loop-1loop-relation} are then determined with the help of IBP relations. This way for example we get
\begin{multline}
J^{(1)}_{111} = 
\frac{1}{z^2}\Big[
\frac{\text{II}_2}{z_1}-\frac{\text{II}_3}{4 z_1}+\frac{\text{II}_4}{2 z_1}-\frac{(2 z-1)
	\text{II}_6}{4 z_2}-\frac{(2 z-1) \text{II}_7}{z_2}
\Big]
+
\frac{\epsilon}{z^2}  \Big[-\frac{16 \text{II}_{\text{1,2}}}{z_1}+\frac{4 \text{II}_{\text{1,3}}}{z_1}-\frac{8
	\text{II}_{\text{1,4}}}{z_1}+12 \text{II}_{\text{2,2}}\\-3 \text{II}_{\text{2,3}}+6
\text{II}_{\text{2,4}}-\frac{3 \text{II}_{\text{2,8}}}{z_1}-3 \text{II}_{\text{3,2}}+\frac{3
	\text{II}_{\text{3,3}}}{4}-\frac{3 \text{II}_{\text{3,4}}}{2}+\frac{3 \text{II}_{\text{3,8}}}{4
	z_1}-\frac{2 (2 z-1) \text{II}_{\text{5,6}}}{z_2}-\frac{8 (2 z-1)
	\text{II}_{\text{5,7}}}{z_2}-\frac{3 \text{II}_{\text{6,6}}}{4}\\-3 \text{II}_{\text{6,7}}+\frac{3 (2
	z-1) \text{II}_{\text{6,8}}}{4 z_2}-3 \text{II}_{\text{7,6}}-12 \text{II}_{\text{7,7}}+\frac{3 (2
	z-1) \text{II}_{\text{7,8}}}{z_2}+\frac{2
	\text{II}_{\text{8,2}}}{z_1}-\frac{\text{II}_{\text{8,3}}}{2
	z_1}+\frac{\text{II}_{\text{8,4}}}{z_1}-\frac{(2 z-1) \text{II}_{\text{8,6}}}{2 z_2}\\-\frac{2 (2 z-1)
	\text{II}_{\text{8,7}}}{z_2}-\frac{\pi ^2 \left(2 z z_1-z_1+z_2\right)}{24 z_1
	z_2}\Big] + \OO (\ep^2)
\end{multline}
and as a consequence of relation \eqref{eq:2loop-1loop-relation} the $J_8$ master integral ($J^{mM}_{01111}$) is given by\footnote{See Appendix \ref{II-notation} for $\text{J}_{\ldots}$ iterated integrals notation.} 
\begin{multline}
J_8 = 
\frac{\text{J}_{\bar{1},6}}{16}+\frac{\text{J}_{\bar{1},7}}{4}-\frac{\text{J}_{\bar{3},6}}{32}-\frac{\text{J}_{\bar{3},7}}{8}-\frac{\text{J}_{\bar{4},2}}{8}+\frac{\text{J}_{\bar{4},3}}{32}-\frac{\text{J}_{\bar{4},4}}{16}
+
\epsilon 
\Big[\frac{\text{J}_{\bar{5},6}}{16}+\frac{\text{J}_{\bar{5},7}}{4}-\frac{\text{J}_{\bar{6},6}}{32}-
\frac{\text{J}_{\bar{6},7}}{8}-\frac{\text{J}_{\bar{7},2}}{8}+\frac{\text{J}_{\bar{7},3}}{32}-\frac{\text{J}_{\bar{7},4}}{16}+\frac{1}{2} \text{J}_{\bar{1},5,6}\\+2 \text{J}_{\bar{1},5,7}-\frac{3}{16}
\text{J}_{\bar{1},6,8}-\frac{3}{4} \text{J}_{\bar{1},7,8}+\frac{1}{8}
\text{J}_{\bar{1},8,6}+\frac{1}{2} \text{J}_{\bar{1},8,7}-\frac{3}{2}
\text{J}_{\bar{2},2,2}+\frac{3}{8} \text{J}_{\bar{2},2,3}-\frac{3}{4}
\text{J}_{\bar{2},2,4}+\frac{3}{8} \text{J}_{\bar{2},3,2}-\frac{3}{32}
\text{J}_{\bar{2},3,3}+\frac{3}{16} \text{J}_{\bar{2},3,4}\\+\frac{3}{32}
\text{J}_{\bar{2},6,6}+\frac{3}{8} \text{J}_{\bar{2},6,7}+\frac{3}{8}
\text{J}_{\bar{2},7,6}+\frac{3}{2} \text{J}_{\bar{2},7,7}-\frac{1}{4}
\text{J}_{\bar{3},5,6}-\text{J}_{\bar{3},5,7}+\frac{3}{32} \text{J}_{\bar{3},6,8}+\frac{3}{8}
\text{J}_{\bar{3},7,8}-\frac{1}{16} \text{J}_{\bar{3},8,6}-\frac{1}{4} \text{J}_{\bar{3},8,7}+2
\text{J}_{\bar{4},1,2}\\-\frac{1}{2} \text{J}_{\bar{4},1,3}+\text{J}_{\bar{4},1,4}+\frac{3}{8}
\text{J}_{\bar{4},2,8}-\frac{3}{32} \text{J}_{\bar{4},3,8}-\frac{1}{4}
\text{J}_{\bar{4},8,2}+\frac{1}{16} \text{J}_{\bar{4},8,3}-\frac{1}{8}
\text{J}_{\bar{4},8,4}-\frac{1}{8} \log (2) \text{J}_{\bar{1},6}-\frac{1}{2} \log (2)
\text{J}_{\bar{1},7}+\frac{1}{16} \log (2) \text{J}_{\bar{3},6}\\+\frac{1}{4} \log (2)
\text{J}_{\bar{3},7}+\frac{1}{4} \log (2) \text{J}_{\bar{4},2}-\frac{1}{16} \log (2)
\text{J}_{\bar{4},3}+\frac{1}{8} \log (2) \text{J}_{\bar{4},4}+\frac{1}{96} \pi ^2
\text{J}_{\bar{1}}-\frac{1}{192} \pi ^2 \text{J}_{\bar{3}}+\frac{1}{192} \pi ^2
\text{J}_{\bar{4}}\Big] + \OO (\ep^2)
\end{multline}
Similarly one can get expressions for all other two-loop master integrals at any desired order in $\ep$-expansion.

\section{Integral representations for general values of $d$}
\label{exactIntegralReps}

Now let us consider how one can get integral representations of our master integrals for general values of space-time dimension. As an example consider the $J^{mM}_{01111}$ master integral. The starting point is the expression for $J^{mM}_{01111}$ master integral in terms of hypergeometric functions in Eq. \eqref{eq:J8-hypergeometry}. Then the problem of obtaining integral representation for this master integral is reduced to the problem of getting integral representations for corresponding hypergeometric functions. The latter problem can be solved using the techniques presented in \cite{KKOVelliptic2,kotikov-EMPL}. First consider ($d=2-2\ep$)
\begin{multline}
F_{01111}^{(1)}(x) = ~_5 F_4 \left(
\begin{array}{c}
1, \frac{1}{2}+\ep, 1+\ep, \frac{1+\ep}{2}, \frac{2+\ep}{2} \\
\frac{3+\ep}{2}, \frac{3}{4}+\frac{\ep}{2}, \frac{5}{4}+\frac{\ep}{2},\frac{1-\ep}{2}
\end{array} \Big| -\frac{x^2}{4}
\right) \\
= \sum_{m=0}^{\infty} \frac{\Gamma (1+m+\frac{\ep}{2})\Gamma (1+2m+2\ep)}{(1+2m+\ep)\Gamma (\frac{1}{2}+m-\frac{\ep}{2})\Gamma (\frac{3}{2}+2m+\ep)} \frac{2^{\ep}\sqrt{\pi}(1+\ep)\Gamma (\frac{3}{2}+\ep)\sec (\frac{\pi\ep}{2})}{\Gamma (1+\ep)\Gamma (1+2\ep)} \left(-\frac{x^2}{4}\right)^m\, .
\end{multline}
Using integral representations for the ratio of $\Gamma$-functions and simple fraction
\begin{align}
\frac{\Gamma (1+2m+2\ep)}{\Gamma (\frac{3}{2}+2m+\ep)} &= \frac{1}{\Gamma (\frac{1}{2}-\ep)}\int_0^1 dp~ p^{2m+2\ep} (1-p)^{-1/2-\ep}\, , \\
\frac{1}{1+2m+\ep} &= \int_0^1 dx_1~ x_1^{2m+\ep}
\end{align}
we get
\begin{align}
F_{01111}^{(1)}(x) &= \frac{\Gamma (\frac{3}{2}+\ep)}{\Gamma (\frac{1}{2}-\ep)\Gamma (1+2\ep)}\int_0^1 dp p^{2\ep} (1-p)^{-\frac{1}{2}-\ep} ~_3 F_2 \left(
\begin{array}{c}
1, \frac{1+\ep}{2}, \frac{2+\ep}{2} \\
\frac{1-\ep}{2}, \frac{3+\ep}{2}
\end{array} \Big| -\frac{p^2x^2}{4} 
\right) \label{eq:F32-rep}\\
&= \frac{(1+\ep)\Gamma (\frac{3}{2}+\ep)}{\Gamma (\frac{1}{2}-\ep)\Gamma (1+2\ep)}
\int_0^1 dp p^{2\ep} (1-p)^{\frac{1}{2}-\ep}
\int_0^1 dx_1 x_1^{\ep} ~_2 F_1 (1,\frac{2+\ep}{2};\frac{1-\ep}{2};-\frac{p^2 x_1^2 x^2}{4})\, .
\end{align}
Next using series representation for $~_2 F_1$-function:
\begin{equation}
~_2 F_1 (1,\frac{2+\ep}{2};\frac{1-\ep}{2};-\frac{p^2 x_1^2 x^2}{4}) = \sum_{m=0}^{\infty} \frac{\Gamma (1+m+\frac{\ep}{2})}{\Gamma (1+m-\frac{\ep}{2})} \frac{\Gamma (\frac{1-\ep}{2})}{\Gamma (\frac{2+\ep}{2})} \left(-\frac{p^2 x_1^2 x^2}{4}\right)^m
\end{equation}
together with integral representation for the ratio of $\Gamma$-functions
\begin{equation}
\frac{\Gamma (1+m+\frac{\ep}{2})}{\Gamma (\frac{1}{2}+m-\frac{\ep}{2})} = \frac{1}{\Gamma (-\frac{1}{2}-\ep)}\int_0^1 dx_2 x_2^{m+\ep/2} (1-x_2)^{-3/2-\ep}
\end{equation}
we finally obtain
\begin{equation}
F_{01111}^{(1)}(x) = \frac{4 (1+\ep)\Gamma (\frac{1-\ep}{2})\Gamma (\frac{3}{2}+\ep)}{\Gamma (-\frac{1}{2}-\ep)\Gamma (\frac{1}{2}-\ep)\Gamma (1+\frac{\ep}{2})\Gamma (1+2\ep)}\int_0^1 dp p^{2\ep} (1-p)^{-\frac{1}{2}-\ep} K^{(1)}(px)\, ,
\end{equation}
where
\begin{equation}
K^{(1)}(px) = \int_0^1 dx_1 x_1^{\ep}\int_0^1 dx_2 \frac{x_2^{\ep/2}(1-x_2)^{-3/2-\ep}}{4+p^2 x_1^2 x_2 x^2}\, .
\end{equation}
To perform $\ep$-expansion of this hypergeometric function and all others which will follow it is convenient to use integral representation\footnote{and similar representations for other hypergeometric functions} \eqref{eq:F32-rep}  and expand $_3 F_2$ function. The latter can be done either using available packages like \texttt{HypExp} \cite{HypExp1,HypExp2} or writing down its differential system and solving it in terms of multiple polylogarithms using reduction to $\ep$-basis. 

Similary for other hypergeometric functions entering expression for $J^{mM}_{01111}$ we have
\begin{align}
F_{01111}^{(2)}(x) &= ~_5 F_4 \left(
\begin{array}{c}
1, 1+\ep, \frac{3}{2}+\ep, 1+\frac{\ep}{2}, \frac{3+\ep}{2} \\
2+\frac{\ep}{2}, \frac{5}{4}+\frac{\ep}{2}, \frac{7}{4}+\frac{\ep}{2}, 1-\frac{\ep}{2}
\end{array} \Big| -\frac{x^2}{4}
\right) \\
&= \frac{\Gamma (\frac{5}{2}+\ep)}{\Gamma (\frac{1}{2}-\ep)\Gamma (2+2\ep)} \int_0^1 dp p^{1+2\ep} (1-p^{-\frac{1}{2}-\ep})
~_3 F_2 \left(
\begin{array}{c}
1, 1+\frac{\ep}{2}, \frac{3+\ep}{2} \\
\frac{2-\ep}{2}, \frac{4+\ep}{2}
\end{array} \Big| -\frac{p^2x^2}{4}
\right) \\
&= \frac{4 (2+\ep)\Gamma (1-\frac{\ep}{2})\Gamma (\frac{5}{2}+\ep)}{\Gamma (-\frac{1}{2}-\ep)\Gamma (\frac{1}{2}-\ep)\Gamma (\frac{3+\ep}{2})\Gamma (2+2\ep)}
\int_0^1 dp p^{1+2\ep}(1-p)^{-1/2-\ep} K^{(2)}(px)\, ,
\end{align}
where
\begin{equation}
K^{(2)}(px) = \int_0^1 dx_1 x_1^{1+\ep}\int_0^1 dx_2 \frac{x_2^{1/2+\ep/2}(1-x_2)^{-3/2-\ep}}{4+p^2x_1^2 x_2 x^2}\, .
\end{equation}

\begin{align}
F_{01111}^{(3)}(x) &= ~_5 F_4 \left(
\begin{array}{c}
1, 1, \frac{3}{2}, 1+\frac{\ep}{2}, \frac{3}{2}+\frac{\ep}{2} \\
2, \frac{5}{4}, \frac{7}{4}, 1-\ep
\end{array} \Big| -\frac{x^2}{4}
\right) \\ 
&= \frac{3\sqrt{\pi}}{4\Gamma (\frac{1}{2}-\ep)\Gamma (2+\ep)}
\int_0^1 dp p^{1+\ep} (1-p)^{-1/2-\ep} ~_3 F_2 \left(
\begin{array}{c}
1, 1, \frac{3}{2} \\
2, 1-\ep
\end{array} \Big| -\frac{p^2x^2}{4}
\right) \\
&= \frac{6\Gamma (1-\ep)}{\Gamma (-\frac{1}{2}-\ep)\Gamma (\frac{1}{2}-\ep)\Gamma (2+\ep)}\int_0^1 dp p^{1+\ep} (1-p)^{-1/2-\ep} K^{(3)}(px)\, ,
\end{align}
where
\begin{equation}
K^{(3)}(px) = \int_0^1 dx_1\int_0^1 dx_2 \frac{x_2^{1/2}(1-x_2)^{-3/2-\ep}}{4+p^2 x_1 x_2 x^2}\, .
\end{equation}

\begin{align}
F_{01111}^{(4)}(x) &= ~_4 F_3 \left(
\begin{array}{c}
\frac{1}{2}, 1, 1+2\ep, 1+\ep \\ 2+\ep, \frac{3}{2}+\ep, -\ep
\end{array} \Big| x
\right) \\
&= \frac{\Gamma (\frac{3}{2}+\ep)}{\Gamma (\frac{1}{2}-\ep)\Gamma (1+2\ep)}\int_0^1 dp p^{2\ep} (1-p)^{-1/2-\ep} ~_3 F_2 \left(
\begin{array}{c}
\frac{1}{2}, 1, 1+\ep \\ -\ep, 2+\ep 
\end{array} \Big| px 
\right) \\
&= \frac{(1+\ep)\Gamma (-\ep)\Gamma (\frac{3}{2}+\ep)}{\sqrt{\pi}\Gamma (-\frac{1}{2}-\ep)\Gamma (\frac{1}{2}-\ep)\Gamma (1+2\ep)}\int_0^1 dp p^{2\ep} (1-p)^{-1/2-\ep} K^{(4)}(px)\, ,
\end{align}
where
\begin{equation}
K^{(4)}(px) = \int_0^1 dx_1 x_1^{\ep}\int_0^1 dx_2 \frac{x_2^{-1/2}(1-x_2)^{-3/2-\ep}}{1-p x_1 x_2 x}
\end{equation}

\begin{align}
F_{01111}^{(5)}(x) &= ~_4 F_3 \left(
\begin{array}{c}
1, 1, 1+\ep, \frac{1}{2}-\ep \\
2, \frac{3}{2}, -2\ep
\end{array} \Big| x
\right) \\
&= \frac{2^{-2(1+\ep)}\Gamma (-\ep)}{\Gamma (-2\ep)\Gamma (1+\ep)}
\int_0^1 dp p^{\ep} (1-p)^{-1/2-\ep} ~_3 F_2 \left(
\begin{array}{c}
1, 1, \frac{1}{2}-\ep \\
2, -2\ep
\end{array} \Big| px
\right) \\
&= \frac{2^{-2 (1+\ep)}\Gamma (-\ep)}{\Gamma (-\frac{1}{2}-\ep)\Gamma (\frac{1}{2}-\ep)\Gamma (1+\ep)}
\int_0^1 dp p^{\ep} (1-p)^{-1/2-\ep} K^{(5)}(px)\, ,
\end{align}
where
\begin{equation}
K^{(5)}(px) = \int_0^1 dx_1\int_0^1 dx_2 \frac{x_2^{-1/2-\ep}(1-x_2)^{-3/2-\ep}}{1-p x x_1x_2}\, .
\end{equation}
The integral expressions for other master integrals can be obtained along the same lines. The main goal of such integral representations is to analytically extend solutions of the type \eqref{eq:J3GeneralSer} valid in the neighborhood of $x=0$ to the whole $x$ plane. Another possible approach to this problem is to look for a series solution in the neighborhood of other critical points $\sum c_n (x - x_c)^n$ including infinity  $\sum c_n/x^n$. Thus, the solution for each region will be expressed by its own, exact in $d$, series, such that different series are matched at the regions of their applicability.
This approach can be especially useful in cases where the solution is expressed in terms of multiple sums.

\section{Conclusion}
\label{conclusion}

In this short note we used an example a set of NRQCD master integrals considered previously in \cite{KKOVelliptic1,KKOVelliptic2,kotikov-EMPL,kalmykov-sunset} to introduce new methods for obtaining  results, which are either exact in the value of space-time dimension or expanded at its fixed value to any prescribed accuracy. The obtained results agree with those obtained previously when either exact results exists or up to available expansion order in $\ep$. The presented techniques for obtaining Frobenius power series and integral representations are both simple and powerful enough with a great potential for their extension to other problems. The presentation in current paper is somewhat sketchy, while more detail exposition will be the subject of one of our future publications.

This work was supported  by  Russian Science Foundation, grant 20-12-00205. The authors also would like to thank Heisenberg-Landau program.

\appendix

\section{Notation for iterated integrals}\label{II-notation}

In this Appendix we define our notation used for iterated integrals with algebraic kernels. Throughout the paper we have two types of such integrals. The first one is of polylogarithmic type, which we define as
\begin{equation}
\text{II}_{a_1\ldots a_n} = \text{II}_{a_1\ldots a_n}(z) = 
\text{II}(\omega_{a_1},\ldots ,\omega_{a_n};z)= \int_0^z \omega_{a_1}(z_1)\int_0^{z_1}\omega_{a_2}(z_2)\ldots \int_0^{z_{n-1}} \omega_{a_n}(z_n)\, ,
\end{equation}
where $\omega_{a_i}$ are algebraic 1-forms. Our results contain only 8 types of such 1-forms, which are given by
\begin{align}
\omega_1(z) &= \frac{z}{z_1^2}dz, & \omega_2(z) &= \frac{z}{z_1}dz, & \omega_3(z) &= \frac{z_1}{z}dz, & \omega_4(z) &= \frac{1}{z_1}dz, & \nonumber \\
\omega_5(z) &= \frac{1}{z_2^2}dz, & \omega_6(z) &= \frac{z_2}{z}dz, & \omega_7(z) &= \frac{1}{z_2}dz, & \omega_8(z) &= \frac{1}{z}dz. & \label{eq:polylog-1forms}
\end{align}
Here are $z_1 = \sqrt{1+4z^2}$ and $z_2 = \sqrt{1-4z}$. In particular we have
\begin{equation}
\text{II}_{1,4} = \int_0^z\frac{z'dz'}{1+4{z'}^2}\int_0^{z'}\frac{dz''}{\sqrt{1+4{z''}^2}}\, ,\quad \text{II}_{3} = \int_0^z\frac{\sqrt{1+4{z'}^2}}{z'}dz'\, .
\end{equation}
The second type of iterated integrals is of elliptic type. We define them as
\begin{multline}
\text{J}_{\bar{a}_1,\ldots ,a_n} = \text{J}_{\bar{a}_1,\ldots ,a_n}(x) = 
\text{J}(\omega_{\bar{a}_1},\ldots ,\omega_{a_n};x) = \int_0^1 \omega_{\bar{a}_1}(t) \text{II}(\omega_{a_2},\ldots ,\omega_{a_n};\frac{xt}{4})
\\
= \int_0^1 \omega_{\bar{a}_1}(t) \int_0^{\frac{xt}{4}}\omega_{a_2}(z_2)\int_0^{z_2}\omega_{a_3}(z_3)\ldots \int_0^{z_{n-1}}\omega_{a_n}(z_n)
\end{multline}
Here all 1-forms $\omega_{a_i}$ except the first one are the same as in Eq. \eqref{eq:polylog-1forms}. The first 1-forms are different and take the following expressions
\begin{gather}
\omega_{\bar 1}(t) = \Omega_{1,0}^{1,0,1}(t),\quad  \omega_{\bar 2}(t) = \Omega_{1,0}^{2,0,0}(t),\quad  \omega_{\bar 3}(t) = \Omega_{1,0}^{2,0,1}(t),\quad  \omega_{\bar 4}(t) = \Omega_{1,0}^{2,1,0}(t), \nonumber   \\ \omega_{\bar 5}(t) = \Omega_{1,1}^{1,0,1}(t),\quad  \omega_{\bar 6}(t) = \Omega_{1,1}^{2,0,1}(t),\quad \omega_{\bar 7}(t) = \Omega_{1,1}^{2,1,0}(t),
\end{gather}
where
\begin{equation}
\Omega_{de}^{abc}(t) = \frac{t^d\log^e t ~dt}{z^a z_1^b z_2^c}\Big|_{z=\frac{xt}{4}}
\end{equation}
In particular we have 
\begin{align}
\text{J}_{\bar 3} &= \int_0^1 \frac{16 dt}{t x^2 \sqrt{1-tx}}, \quad 
\text{J}_{\bar{5},6} = \int_0^1 \frac{4\log (t) dt}{x\sqrt{1-tx}}\int_0^{\frac{xt}{4}}\frac{\sqrt{1-4z'}}{z'}dz',
\nonumber \\
\text{J}_{\bar{4},3,8} &= \int_0^1\frac{16 dt}{t x^2 \sqrt{1+\frac{t^2x^2}{4}}}\int_0^{\frac{xt}{4}}\frac{\sqrt{1+4{z'}^2}}{z'}dz' \int_0^{z'} \frac{dz''}{z''}\, .
\end{align}
It may turn out that the last integration over $t$ is divergent, such as the integral $\text{J}_{\bar 3}$ in our previous example. Therefore, this definition must be supplemented with an appropriate regularization rule. We define it as follows. If the integral $\int_{0}^1 f(t)dt$ diverges at the lower limit of integration then the function $f$ has an expansion $f(t) = \sum_{i=1}^k a_i t^{-i} + \dots$ and it is natural to define the regularized version of integral with the substitution $f(t)\to f^{\text{reg}}(t)$, where $f^{\text{reg}}(t) =f(t) - \sum_{i=1}^k a_i t^{-i}$. This way  for our previous example we get the following regularization prescription
\begin{equation}
\text{J}_{\bar 3}  \rightarrow \text{J}_{\bar 3}^{\text{reg}} =  \int_0^1 \left[ \frac{16}{t x^2 \sqrt{1-tx}} - \frac{16}{tx^2}\right]dt.
\end{equation}
Note, that the final results for our master integrals do not depend on regularization prescription due to cancellation of divergences between different $\text{J}$-functions.

\bibliographystyle{hieeetr}
\bibliography{litr}

\end{document}